\documentclass[pre,twocolumn,aps,floatfix,10pt]{revtex4-1}
\usepackage{amsmath}
\usepackage{amsfonts}
\usepackage{amssymb}
\usepackage{graphicx}
\usepackage{fontenc}
\usepackage{psfrag,layout}

\begin{document}
\title{Free energy cost of forming an interface between a crystal and its frozen version}
\author{Ronald Benjamin and J{\"u}rgen Horbach}
\affiliation{Institut f{\"u}r Theoretische Physik II, Universit{\"a}t D{\"u}sseldorf,
Universit\"atsstra\ss e 1, 40225 D{\"u}sseldorf, Germany}

\begin{abstract}
Using a thermodynamic integration scheme, we compute the free energy
cost per unit area, $\gamma$, of forming an interface between a crystal
and a frozen structured wall, formed by particles frozen into the same
equilibrium structure as the crystal. Even though the structure and
potential energy of the crystalline phase in the vicinity of the wall
is same as in the bulk, $\gamma$ has a non-zero value and increases with
increasing density of the crystal and the wall. Investigating the effect
of several interaction potentials between the particles, we observe
a positive $\gamma$ at all crystalline densities if the potential is
purely repulsive.  For models with attractive interactions, such as the
Lennard-Jones potential, a negative value for $\gamma$ is obtained at
low densities.
\end{abstract}

\maketitle

\section{Introduction}
\label{sect:introduction}
The excess free energy of a crystal in contact
with a wall is an important quantity governing many
interfacial phenomena such as wetting and heterogeneous
nucleation~\cite{degennes85,zettlemoyer69,abraham74,kashchiev00,adamson97}.
Theoretical approaches based on cell theory have been used
to compute this quantity for model systems such as hard
spheres~\cite{heni-loewen99}.  Atomistic simulation studies
have also been carried out employing widely-used interaction
potentials such as Lennard-Jones (LJ) models and systems of hard
spheres~\cite{heni-loewen99,benjamin-horbach2012,benjamin-horbach2013_1,benjamin-horbach2013_2,fort-djik06,laird07}.

In most previous studies, the crystal is either in contact with a
flat structureless wall~\cite{heni-loewen99,fort-djik06,laird07} or a
structured wall consisting of particles attached to sites corresponding
to an ideal crystalline structure~\cite{benjamin-horbach2012}. In both
of these situations, the structure and energy of the crystalline layers
in the neighborhood of the wall is not the same as in the bulk leading
to an excess free energy as compared to the bulk.

An interesting situation arises when the energy and structure of the
crystalline layers in contact with the wall remains the same as in the
bulk. Such a scenario occurs when the wall consists of particles frozen
into positions occupied by the same crystalline phase at equilibrium. In
this case, the wall is a frozen version of the same crystal. Then, it is
of interest from a fundamental statistical mechanical point of view, to
determine the excess free energy of the system in comparison to the bulk.

An analogous situation corresponding to a supercooled liquid
in contact with a wall consisting of particles frozen into the
same amorphous structure as the liquid has been widely studied
in recent years in order to investigate the relaxation dynamics of
glass-forming systems~\cite{krackoviak,scheidler2004,scheidler_epl2002,
berthier2012,cavagna2012,gradenigo2013,kobnaturephys,cammarota2012}.
Recently, an experimental investigation of such a system was carried
out using colloidal particles~\cite{sood2015}.  Such computational and
experimental studies extract a growing length scale based on a slowing
down of the dynamics near the wall and relate it to the slowing down of
glassy dynamics in the bulk.  In these investigations, it is implicitly
assumed that the thermodynamics of the system is unperturbed by the wall
if an average is carried out over the thermal fluctuations and different
realizations of the wall~\cite{scheidler2004,berthier2012}. However,
in a recent work we obtained a non-zero interfacial free energy for a
glass-forming binary Lennard-Jones liquid~\cite{kob-andersen-1995} in
contact with a wall with the same amorphous structure as the supercooled
liquid, indicating that thermodynamics of the liquid is affected by the
presence of such a frozen wall~\cite{benjamin-horbachpre}.

In this work, we compute the excess free energy of a crystal in
contact with a wall formed by its frozen version, using molecular
dynamics simulations~\cite{allen-tildesley87} in conjunction with
thermodynamic integration~\cite{frenkel-smit02}.  We obtained a non-zero
interfacial free energy for particles interacting via a Lennard-Jones
potential~\cite{benjamin-horbach2014}, which increases with increasing
density of the crystal suggesting that the thermodynamics of the crystal
in contact with such a wall is perturbed and not the same as in the
bulk. Moreover, at low densities $\gamma$ is negative, indicating an
increase of entropy as compared to the bulk.  To test the robustness
of our results and find out whether a negative value of $\gamma$
is model-dependent or not, we carried out investigations with two
other interaction potentials: a purely repulsive inverse-twelfth power
soft-sphere potential and the hard sphere interaction (using a $r^{\rm
-256}$ potential)~\cite{benjamin-horbach2015}.  However, $\gamma$ for
the two purely repulsive interactions, while non-zero, is positive  at
all crystal densities.

In the next section we specify the interaction potentials considered
in this work. In section~\ref{sect:system_setup}, the system is
described and then we outline the thermodynamic integration scheme
adopted to compute $\gamma$. In section~\ref{sect:simulation},
we describe the simulation details and then present our results in
section~\ref{sect:results}. Finally, we end with a conclusion.

\section{Interaction Potentials} 
\label{sect:potentials}
In order to study the robustness of our findings, we obtained
$\gamma$ corresponding to three different interaction potentials. The
interaction potentials are denoted by $u_{\rm i}(r)$ [$i=1,2,3$],
$r$ being the distance between the particles.  Energies and lengths
are expressed in units of the parameters $\varepsilon$ and $\sigma$,
respectively (see below), and the masses of the particles are set to
$m=1$. In the following, all other quantities are given in terms of the
latter parameters, i.e.~time $t$ in units of $\tau = \sqrt{m\sigma^{\rm
2}/\varepsilon}$, pressure $P$ in units of $\varepsilon/\sigma^{\rm 3}$,
and the interfacial free energy in units of $k_{\rm B}T/\sigma^{\rm 2}$.

The following interaction potentials are considered in this work:

(i) A modified Lennard-Jones interaction potential (mLJ) is chosen,
defined by
\begin{equation}
u_{1}(r_{ij}) = \phi(r_{ij}) + C_{1} 
\label{eq:ur1}
\end{equation}
with
\begin{equation}
\phi(r_{ij}) =
4\epsilon 
\left[\left(\frac{\sigma}{r_{ij}}\right)^{12}
- \left(\frac{\sigma}{r_{ij}}\right)^{6} \right],
\label{eq:lj}
\end{equation}
for $0<r_{ij}\leq2.3\,\sigma$, and
\begin{equation}
 u_{1}(r_{ij}) = C_{2} \left(\frac{\sigma}{r_{ij}}\right)^{12} 
+ C_{3} \left(\frac{\sigma}{r_{ij}}\right)^{6}
+ C_{4}\left(\frac{r_{ij}}{\sigma}\right)^{2} + C_{5} 
\label{eq:ur2}
\end{equation}
for $2.3\,\sigma<r_{ij}<r^{\rm cut}=2.5\,\sigma$ and $u(r_{ij})=0$
for $r_{ij}\geq r^{\rm cut}$. The constants in Eqs.~(\ref{eq:ur1})
and (\ref{eq:ur2}) are given by $C_{1}=0.016132\,\epsilon$,
$C_{2}=3136.6\,\epsilon$, $C_{3}=-68.069\,\epsilon$,
$C_{4}=-0.083312\,\epsilon$, and $C_{5}=0.74689\,\epsilon$.

(ii) The second potential we consider is the purely repulsive
force-shifted inverse-twelfth power potential (fsi12),
\begin{equation}
u_{2}(r_{ij}) = \psi(r_{\rm ij})-\psi(r_{\rm c})
-(r-r_{\rm c})\left.\frac{d \psi(r_{\rm ij})}
{d r_{\rm ij}}\right|_{r_{\rm ij}=r_{\rm c}}
\end{equation}
where, $\psi(r_{\rm ij})=4.0 \epsilon (\sigma/r_{\rm ij})^{\rm 12}$.
Here, the potential is truncated and shifted such that $u_{2}(r_{ij})$ and
its derivative with respect to $r_{ij}$ goes to zero at $r_{c}=2.5\sigma$.

(iii) The third potential is a hard-sphere interaction modeled by a
inverse-power potential (i256):
\begin{equation}
u_{3}(r_{ij}) =
\epsilon
\left(\frac{\sigma}{r_{ij}}\right)^{256}\;.
\label{eq:pot2}
\end{equation}
The potential is cut off at $r_{c}=1.2\sigma$. As shown previously
\cite{benjamin-horbach2015}, the coexistence and other bulk properties of
this interaction potential are very similar to those of the hard-sphere
system.

\begin{figure}[tba]
\includegraphics[width=3.0in]{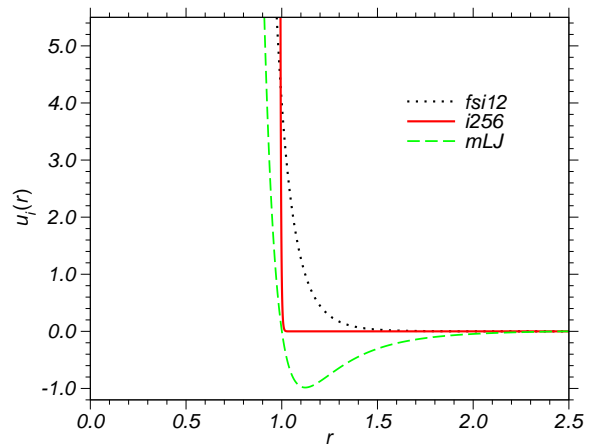}
\caption{\label{fig1}(Color online) Interaction potentials corresponding
to a modified LJ potential (mLJ), the inverse-256th power interaction
approximating the hard sphere potential (i256), and the force-shifted
inverse-twelfth power potential (fsi12).}
\end{figure}
In Fig.~\ref{fig1}, we show the three interaction potentials as a function
of the distance $r$ between the particles.

\section{System Setup}
\label{sect:system_setup}
\begin{figure}[tba]
\includegraphics[width=3.0in]{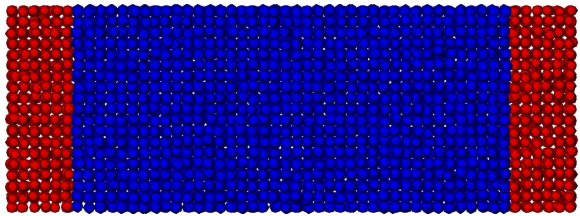}
\caption{\label{fig2}(Color online) Crystal (blue) in contact with
several layers of frozen-in crystalline particles (red), forming walls
on either side. This snapshot represents an instataneous state of
the system at equilibrium corresponding to the modified Lennard-Jones
potential (mLJ) at $\rho=1.1$ and $T=1.0$.}
\end{figure}

Our systems consist of face centered cubic (fcc) crystals of $N$
particles, with the (100) face oriented along the $z$ direction, enclosed
in a simulation box of dimension $L_{\text{x}} \times L_{\text{y}}
\times L_{\text{z}}$. Periodic boundary conditions are maintained along
the $x$ and $y$ directions.  Along the $z$ direction, the particles are
confined by several layers of the same crystalline particles frozen into
their equilibrium configuration (see Fig.~\ref{fig2}).  Since there
are two wall-crystal interfaces, the total area of the interface is
$A=2L_{\text{x}}L_{\text{y}}$.  In order to prevent any crystalline
particle from penetrating the wall, a short-ranged flat wall, modeled
by a Gaussian potential is inserted at the boundaries along the $z$
direction (at $z=0$ and $L_{\rm z}$). As shown in our earlier works
\cite{benjamin-horbach2014,benjamin-horbach2015}, such a short-ranged
flat wall leads to a negligible contribution to the interfacial free
energy ($\approx 0.001 \gamma$).

\section{Thermodynamic Integration} 
\label{sect:TI}
The interfacial free energy is defined as $\gamma=(F_{\rm system}
-F_{\rm bulk})/A$, where $F_{\rm system}$ is the free energy of the
crystal confined by the frozen walls on both sides, while $F_{\rm bulk}$
is the free-energy of the bulk crystal with periodic boundary conditions
along all the three Cartesian axes. Our objective is to compute $\gamma$
using molecular dynamics simulation and thermodynamic integration.

The thermodynamic integration scheme adopted in this work is based on
a similar approach devised by us in previous studies to compute the
interfacial free energy of a liquid/crystal in contact with structured
walls~\cite{benjamin-horbach2012,benjamin-horbach2013_1,benjamin-horbach2013_2}
and to determine the crystal-liquid interfacial free
energy~\cite{benjamin-horbach2014,benjamin-horbach2015}.  The goal
is to transform a bulk crystal with periodic boundaries in all three
Cartesian-axes directions into a crystal in contact with frozen
crystalline layers comprising the frozen walls on either side. To
accomplish this, it is neccessary that the boundary conditions
along the $z$ direction are rearranged at some point during the
transformation and that the crystalline particles do not cross the
boundaries during this rearrangement. Hence, in the first step, two
extremely short-ranged flat walls, modeled by a Gaussian potential
and denoted by $u_{\text{fw}}(z^\prime)$, are gradually inserted at
the boundaries of the simulation box at $z=0$ and $z=L_{\rm z}$ with
periodic boundary conditions maintained along all three axes. Since
the structureless flat wall has a very short range ($0.001\sigma$), few
crystalline particles interact with it. As a result, the contribution
of this step to the total interfacial free energy is negligible and does
not affect the structure of the crystal near the walls. To integrate the
equations of motion in molecular dynamics in presence of such extremely
short ranged walls, a very small time-step is needed. This reduces
the computational efficiency since the other forces in the simulation
such as those between the crystalline particles are more comparatively
long-ranged ($\sim \sigma$). However, by the use of a multiple time-step
scheme~\cite{frenkel-smit02,benjamin-horbach2014,benjamin-horbach2015},
the computational overhead is reduced significantly and our simulations
are slightly less than two times slower than those carried out with a
single large time-step.

In the second step, interactions of particles through the periodic
boundaries along the $z$ direction are gradually switched off while
interactions between the crystal and the frozen crystalline wall are
gradually switched on, in presence of the flat walls.  The final state is
a crystal with periodic boundaries along the $x$ and $y$ axes and confined
by frozen crystalline walls and the flat walls along the $z$ direction.
Since the free-energy difference per unit area corresponding to the
first step is negligible the total contribution to $\gamma$ comes only
from the second step.

In the TI scheme adopted in this work, similar to our earlier works, the
transformation from one state to the next is brought about by changing
a switching parameter $\lambda$, which couples to the interaction
potential between the walls and the crystalline particles. The TI
scheme consists of two steps. In step 1, the $\lambda$ dependent
Hamiltonian has the form $H_{1}(\lambda)= \lambda^{2} u_{\rm fw}(z)$,
where $u_{\rm fw}(z)=a \exp(-[(z-z_{\rm w})/b]^{\rm 2})$, with $a=25.0
k_{\rm B}T$ and $b=0.001\sigma$.  By switching $\lambda$ from $0$ to
$1$, the system is transformed from a pure bulk crystal ($\lambda=0$, no
wall), to a crystal in contact with an extremely short-ranged Gaussian
flat wall ($\lambda=1$).  In step 2, the $\lambda$-dependent part of
the Hamiltonian has different forms for the different interaction
potentials.  For the LJ and inverse-twelfth power potentials,
$H_{2}(\lambda)=(1-\lambda)^{2}u^{\ast}(r)+ \lambda^{2}u_{w}(r)$, while
for the hard-sphere potential, $H_{2}(\lambda)=(1-\lambda)^{8}u^{\ast}(r)+
\lambda^{256}u_{w}(r)$. Here, $u^{\ast}(r)$ represents interactions
between the crystalline particles through the periodic boundaries,
while $u_w$ denotes the interaction between a crystal and a frozen wall
particle and is identical to the interaction between two crystalline
particles. In the second step, $\lambda=0$ corresponds to the crystal
in contact only with a flat wall, while at $\lambda=1$ the crystal is
in contact with the frozen wall in presence of the short-ranged flat wall.

The free-energy difference in each step is computed from the relation
\begin{equation}
\Delta F_{\rm i} = \int_{\lambda = {\rm 0}}^{\lambda = {\rm 1}} 
\left \langle \frac{\partial H_{\rm i}}{\partial \lambda} 
\right \rangle  d\lambda \, ,
\end{equation}
where the integration over $\lambda$ is computed using Simpson's rule.
Finally, the interfacial free-energy is directly obtained from $\Delta
F_1$ and $\Delta F_2$ via $\gamma=({\Delta F_{1}+\Delta F_{2}})/{A}$.

\begin{figure}[tba]
\includegraphics[width=3.0in]{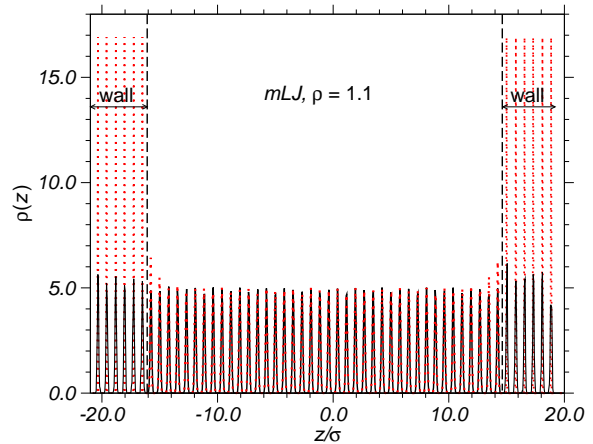}
\caption{\label{fig3}(Color online) Density profile of crystal in contact
with a wall comprising frozen-in crystalline layers (solid line) and a
wall consisting of particles arranged in an ideal fcc lattice structure
having the same density as the crystal (dotted line). The density profiles
were averaged over $1000$ independent configurations of the crystal in
contact with the same frozen wall. The vertical dashed lines indicate the
positions of the short-ranged flat walls at $z=0$ and $L_{\rm z}$. The
profiles were obtained for the mLJ potential at the density $\rho=1.1$ 
and the temperature $T=1.0$.}
\end{figure}
\section{Simulation}
\label{sect:simulation}
To integrate the equations of motion, the velocity form of the Verlet
algorithm~\cite{frenkel-smit02,allen-tildesley87} was implemented
with a multiple time-step scheme. For the mLJ and fsi12 potentials a
smaller time-step $\Delta t_{\rm small} = 0.00025 \tau$ was used along
with a larger time-step $\Delta t_{\rm large} = 0.004 \tau$, while for
the hard-sphere potential, $\Delta t_{\rm small} = 0.00025 \tau$ and
$\Delta t_{\rm large} = 0.0005 \tau$ were chosen. To maintain constant
temperature, $T=1.0$, every 200 time steps the velocity of the particles was
drawn from a Maxwell-Boltzmann distribution at the desired temperature.

We carried out Molecular Dynamics simulations in the $NVT$ ensemble at
various densities of the crystal at the temperature $T=1.0$.  Initially, an ideal fcc crystal was
put in the simulation box and it was ensured that the center of mass of
the crystal along the $z$ direction was at $-L_{\rm z}/2$, i.e.~the two
outermost crystalline layers are at the same distance from the boundaries
at the two ends. At all densities, the crystal was first equilibrated
at the desired temperature in order to chose the reference configurations
for the pinned layers of crystalline particles forming the wall.

The frozen wall is constructed by choosing particle positions from
crystalline slabs of width $2.5 \sigma$ ($1.0 \sigma$ in case of the
hard-sphere potential) from the boundaries at $z=0$ and $z=L_{\rm z}$,
at the two ends of the simulation box, from an equilibrium configuration.
When equilibrating the crystal, every time the thermostat was imposed
the average center of mass momentum of the crystal was set to zero.
New particles fixed at these instantaneous equilibrated positions were
then juxtaposed on the right and left sides of the simulation cell from
the slabs at the left and right ends of width $2.5\sigma$ or $\sigma$
[equal to the cut-off range of the interaction potential $u(r)$],
by shifting their $z$-positions by $-L_{\rm z}$ and $L_{\rm z}$,
respectively.

The system sizes corresponding to the mLJ and fsi12 interaction potentials
consisted of $N=6480$ particles while for the hard-sphere potential,
$N=3600$, were chosen. Simulations carried out at larger system sizes yielded values
of $\gamma$ in agreement with the smaller systems within the statistical
errors. For each interaction potential, $\gamma$ was computed at various
crystal densities ranging from $\rho=1.0-2.0$. At each density,
$\gamma$ was calculated for several realizations of wall configurations.
However, data corresponding to the different realizations were in close
agreement with each other, within the numerical errors. In all values of $\gamma$
reported in the next section, the error bars are less than the symbol size
and hence they are not specified.

\begin{figure}
\includegraphics[width=3.0in]{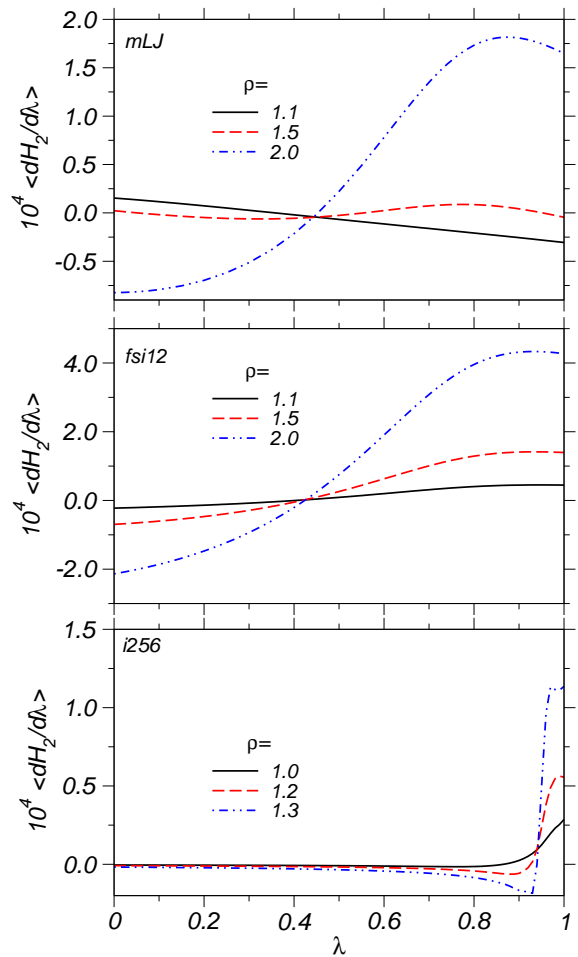}
\caption{\label{fig4}(Color online) Thermodynamic integrand as a function
of $\lambda$ for the second step of the TI scheme at various densities
$\rho$ for (a) the mLJ potential, (b) the fsi12 potential and
(c) the i256 potential.}
\end{figure}
\begin{figure}
\includegraphics[width=3.0in]{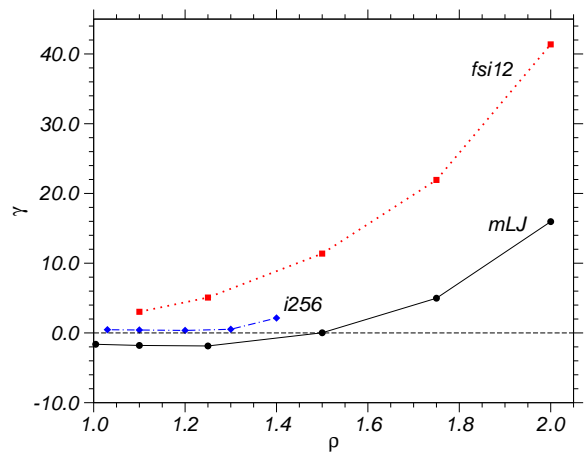}
\caption{\label{fig5}(Color online) Interfacial free energy vs.~density
for the (100) orientation of the crystal in contact with a frozen wall
for the three interaction potentials.  The dashed horizontal line indicates
the zero axis.}
\end{figure}
\section{Results}
\label{sect:results}
Figure~\ref{fig2} shows the simulation set-up of the system with
the crystal in contact with several frozen-in layers of crystalline
particles forming walls on both sides. It is clear from Fig.~\ref{fig2}
that the structure of the crystal near the wall is identical to that
in the bulk.  This is also clear from the density profile shown in
Fig.~\ref{fig3}. There is no change in the density profiles in the
layers in contact with the wall and those far away from it.  We have
also observed that the energy profile of the crystalline layers near the
wall is the same as in the bulk indicating that the wall particles do not
change the structure and energy of the crystal layers in contact with it.
If, instead, the wall is constituted by particles attached to ideal
fcc lattice sites, the peaks of the density  profiles near the wall are
higher than those in layers away from the walls (cf. Fig.~\ref{fig3}),
indicating that the structure of the crystalline layers in the vicinity
of such an ideal wall is different from the bulk.

The thermodynamic integrands corresponding to step 2 of the TI scheme
and pertaining to the different densities are plotted as a function of
$\lambda$ in Fig.~\ref{fig4}, for the three interaction potentials. The
contribution of the Gaussian flat walls in the first step is negligible
(less than $0.1\%$ of $\gamma$) and hence not reported.  A negligible
value of $\Delta F_{\rm 1}$ indicates that the Gaussian flat walls
effectively do not perturb the thermodynamics of the system. The
contribution to $\gamma$ comes almost entirely from the second step.
As Fig.~\ref{fig4} shows, the thermodynamic integrands corresponding to
the two relatively long-ranged interaction potentials are similar, while
that of the short-ranged hard-sphere potential is qualitatively different.
The smoothness of the integrands leads to an accurate determination of
the free-energy difference from numerical integration.

In Fig.~\ref{fig5}, we show $\gamma$ for the three interaction potentials
as a function of the density. For the hard-sphere system, the maximum
density we consider is $\rho=1.4$ since the packing fraction at this
density is $\phi=0.733$, which is very close to the maximum theoretical
packing fraction $\phi_{\rm max}\approx 0.74$. For the other two soft
sphere potentials, we are able to consider larger densities, up to
$\rho=2.0$.

Figure~\ref{fig5} shows a non-zero value for $\gamma$ as a function
of $\rho$ for all the three interaction potentials. The Helmholtz
free-energy $F$ is given by $F=U-TS$, where $U$ is the energy of the
system, $T$ the temperature and $S$ the entropy. Since the potential
energy of the system in contact with the wall is the same as in the
bulk and the system is maintained at the same temperature throughout
the thermodynamic transformation from the initial to the final states,
a non-zero free-energy difference can only be attributed to a change in
the entropy of the system.  A non-zero free energy difference indicates
that the frozen-in layer of crystalline particles imposes an external
field on the crystalline particle, perturbing the thermodynamics as
compared to the bulk.

For the purely repulsive inverse-twelfth power potential and the
hard-sphere potential, the interfacial free energy is always positive
from the lowest to the highest densities considered. As can be seen in
Fig.~\ref{fig1}, for $r/\sigma>1.0$, the inverse twelfth-power potential
is more repulsive as compared to the hard sphere interaction, resulting
in a larger interfacial free energy.  The positive excess free energy
corresponding to both the purely repulsive potentials at all densities and
the modified LJ potential at high densities originates due to a decrease
of the entropy of the particles near the wall as they have less number
of ways to arrange themselves near the wall and the thermal oscillations
of the particles around their mean positions is also suppressed in the
vicinity of the wall.

In case of the modified LJ potential, the interfacial free energy is
positive only at high densities.  As we go to lower and lower densities,
$\gamma$ decreases and eventually becomes negative.  We have also computed
the excess free energy for other interaction potentials and observed
that this negative excess free energy occurs only for potentials with
attractive interactions. A negative $\gamma$ for the mLJ potential at
low densities indicates that the wall exerts an effective attractive
pinning field on the free particles. Since all the contribution to
the change in free-energy comes from the change in entropy $\Delta
S$, a negative $\gamma$ corresponds to $\Delta S >0$, meaning that the
entropy of the crystal in contact with a wall which is its frozen version,
increases at low densities, if the interaction potential has an attractive
component.  As shown in a recent work~\cite{benjamin-horbachpre}, this
counterintuitive finding is similar to the case of a supercooled binary LJ
liquid in contact with an amorphous frozen wall having the same structure
as the liquid, where a negative $\gamma$ was obtained at low temperatures.

\section{Conclusion}
\label{sect:conclusion}
In this work, we computed the excess free energy of a crystal in contact
with a wall composed of several frozen-in layers of the same crystal,
as a function of the density of the crystal. Even though the structure
and energy of the crystal in the vicinity of such a wall is the same as
in the bulk a non-zero interfacial free energy is obtained, whose origin
is purely entropic. A non-zero interfacial free energy indicates that
the wall effectively imposes an external field on the bulk system and
thereby the thermodynamics of the system is altered.

For purely repulsive interaction potentials, the excess free energy is
always positive indicating that entropy of the crystal in contact with
the wall is less than that of a bulk crystal. However, for a modified
LJ potential, which has an attractive component, the interfacial
free energy is negative at low densities, indicating that entropy of the
system in contact with the wall is larger than that of the bulk crystal.

\begin{acknowledgments}
The authors acknowledge financial support from the Deutsche
Forschungsgemeinschaft (DFG) in the framework of the M-era.Net project
``ANPHASES'', grant No.~HO 2231/12-1.
\end{acknowledgments}

\end{document}